\documentclass[aps,prb,twocolumn,superscriptaddress,showpacs]{revtex4}
\usepackage{amssymb}
\usepackage{amsfonts}
\usepackage{xcolor}
\usepackage{amsmath}
\usepackage{graphicx}
\usepackage{bm}

\begin{document}

\title{Charge equilibration in integer and fractional quantum Hall edge channels in a generalized Hall-bar device
}
\author{Chaojing Lin}
\email{lin.c.ad@m.titech.ac.jp}
\author{Ryota Eguchi}
\affiliation{Department of Physics, Tokyo Institute of Technology, 2-12-1 Ookayama,
Meguro, Tokyo, 152-8551, Japan.}
\author{Masayuki Hashisaka}
\affiliation{Department of Physics, Tokyo Institute of Technology, 2-12-1 Ookayama,
Meguro, Tokyo, 152-8551, Japan.}
\affiliation{NTT Basic Research Laboratories, NTT Corporation, 3-1 Morinosato-Wakamiya,
Atsugi, 243-0198, Japan.}
\author{Takafumi Akiho}
\affiliation{NTT Basic Research Laboratories, NTT Corporation, 3-1 Morinosato-Wakamiya,
Atsugi, 243-0198, Japan.}

\author{Koji Muraki}
\affiliation{NTT Basic Research Laboratories, NTT Corporation, 3-1 Morinosato-Wakamiya,
Atsugi, 243-0198, Japan.}
\author{Toshimasa Fujisawa}
\affiliation{Department of Physics, Tokyo Institute of Technology, 2-12-1 Ookayama,
Meguro, Tokyo, 152-8551, Japan.}

\begin{abstract}
Charge equilibration between quantum-Hall edge states can be studied to reveal geometric structure of edge channels not only in the integer quantum Hall (IQH) regime but also in the fractional quantum Hall (FQH) regime particularly for hole-conjugate states. Here we report on a systematic study of charge equilibration in both IQH and FQH regimes by using a generalized Hall bar, in which a quantum Hall state is nested in another quantum Hall state with different Landau filling factors. This provides a feasible way to evaluate equilibration in various conditions even in the presence of scattering in the bulk region. The validity of the analysis is tested in the IQH regime by confirming consistency with previous works. In the FQH regime, we find that the equilibration length for counter-propagating $\delta \nu $ = 1 and $\delta \nu $ = -1/3 channels along a hole-conjugate state at Landau filling factor $\nu $ = 2/3 is much shorter than that for co-propagating $\delta \nu $ = 1 and $\delta \nu $ = 1/3 channels along a particle state at $\nu $ = 4/3. The difference can be associated to the distinct geometric structures of the edge channels. Our analysis with generalized Hall bar devices would be useful in studying edge equilibration and edge structures.

\end{abstract}

\maketitle

\section{Introduction}

Edge channels formed along a boundary of a quantum Hall (QH) system dominate the transport characteristics in integer quantum Hall (IQH) and fractional quantum Hall (FQH) regimes \cite{Halperin82,Macdonal84,Wen90a,Wen90b}. Carrier scattering between edge channels may involve charge transfer between the channels. While the scattering without charge transfer has been recently discussed with energy transfer and Tomonaga-Luttinger physics \cite{ Roulleau08,Sueur10,Altimiras12,Bocquillon13,Tewari16,Itoh18,Ota19}, the scattering with charge transfer is a long-standing crucial problem in justifying the edge channel picture. Charge transfer is basically prohibited in an ideal system, but is actually allowed in the presence of impurity potential. Well defined quantized Hall conductance is associated with significantly suppressed scattering between counter-propagating edge channels along opposite sides of a large QH system \cite{Klitzing80,Tsui82}. If multiple edge channels are formed along one side of the QH system, scattering between them equilibrates the channels, which is referred to as edge equilibration \cite{Landaur70,Buttiker88}. Edge equilibration in the IQH regime is well understood with the edge potential profile for co-propagating edge channels, where the spatial distance between the channels determines the degree of equilibration \cite{Martin90}.

In contrast, edge equilibration in the FQH regime remains veiled \cite{Chang03,Chamon97}. For example, a FQH state at Landau filling factor $\nu $ = 2/3 is believed to have a complex edge structure, where the local filling factor changes non-monotonically from 0 to 1 across an integer edge state with the difference $\delta \nu $ = 1 and then back to 2/3 across a fractional edge state with the difference $\delta \nu $ = -1/3, as proposed by MacDonald \cite{Macdonal90}. This suggests that edge equilibration takes place between counter-propagating channels for integer particles ($\delta \nu $ = 1) and fractional holes ($\delta \nu $ = -1/3). While such edge channels had not been identified for a long time possibly due to full equilibration with significant scattering \cite{Wang13,Kane94}, Grivnin et al., recently reported a feature of edge equilibration between integer (1) and fractional (-1/3) channels for 2/3 state in a short FQH system \cite{Grivnin14}. The measurement relies on the absence of scattering in the bulk FQH state, which can be justified by investigating temperature dependence of thermally activated bulk transport in a high mobility sample. If the scattering in the bulk, which is referred to as bulk equilibration, exists, it can disturb the analysis of edge equilibration. Therefore, it would be better to evaluate both edge and bulk equilibration simultaneously to confirm the absence of bulk equilibration as well as to study the edge equilibration at various conditions even in the presence of bulk equilibration. Moreover, the FQH regime provides a diversity of edge structures \cite{Haldane83,Macdonal90,Kane95,Sabo17}. For example, a quantum Hall state at $\nu $ = 4/3 is expected to have co-propagating channels for integer particles ($\delta \nu $ = 1) and fractional particles ($\delta \nu $ = 1/3). The edge equilibration can be studied systematically for various cases, which might clarify the roles of particles and holes in the FQH physics as well as numerous equilibration phenomena.

In this paper, we study charge equilibration associated with inter-channel charge transfer by employing a generalized Hall bar, in which a Hall-bar shaped circulating edge channels are formed between an inner quantum Hall state and an outer quantum Hall state with different filling factors. The inner and outer states are defined with a Hall-bar shaped gate and a uniform magnetic field. This allows us to address equilibration problems in various IQH and FQH states, especially for the enigmatic hole-conjugate states. First, our analysis on both edge and bulk equilibration is tested in the IQH regime, and we find consistency with previous works. Next, we investigate edge and bulk equilibration in the FQH regime. We find that the length for edge equilibration is significantly shorter for counter-propagating $\delta \nu $ = 1 and $\delta \nu $ = -1/3 channels at $\nu $ = 2/3 as compared to co-propagating $\delta \nu $ = 1 and $\delta \nu $ = 1/3 channels at $\nu $ = 4/3. The scheme can be applied to study edge equilibration for investigating edge structures of hole-conjugate FQH states.

\section{Equilibration in a generalized Hall bar}
\begin{figure}[t]
\includegraphics[width=8 cm]{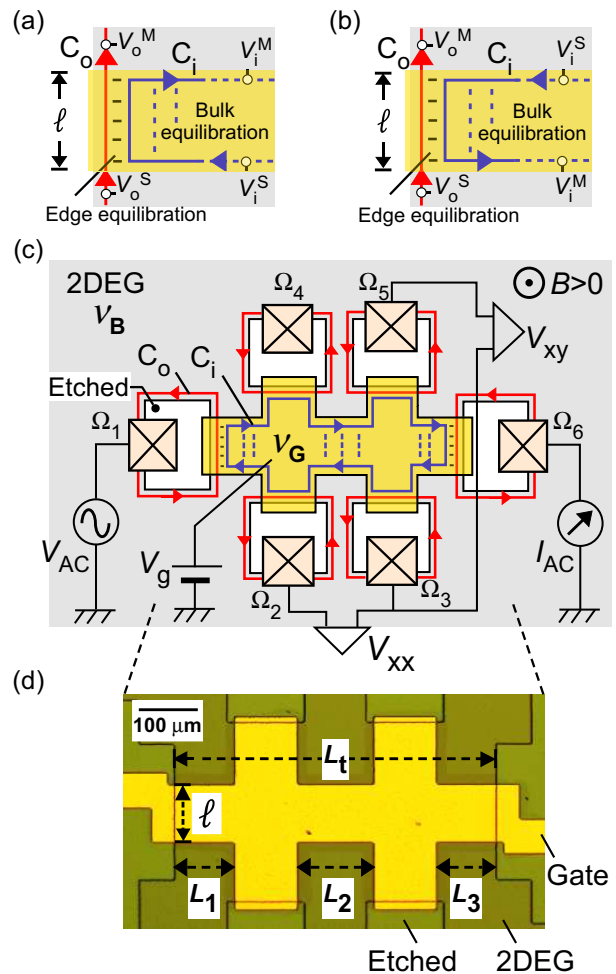}
\caption{(color online). Edge equilibration between (a) co-propagating, and (b) counter-propagating inner and outer channels, $C_{\mathrm{i}}$ and $C_{\mathrm{o}}$, in an interaction region of length $\ell$. Bulk equilibration between the right- and left-going part of $C_{\mathrm{i}}$ may coexist with edge equilibration. (c) Schematic sketch of a generalized Hall bar, which is composed of 6 quasi-Corbino type ohmic contacts $\Omega_{1}$-$\Omega_{6}$ with etched trenches and a Hall bar shaped metal gate. The drawn edge channels are formed at $\nu_{\mathrm{G}}$(= 2) $>$ $\nu_{\mathrm{B}}$(= 1), where closed edge channel(s) $C_{\mathrm{i}}$ formed underneath the gate is coupled to the outer channel(s) $C_{\mathrm{o}}$ by edge equilibration. The measurement setup is also drawn in the figure. (d) An optical image for the central part of the device. Relevant length scales for the metal gate are labeled.}
\end{figure}

\begin{figure*}[t]
\includegraphics[width=16 cm]{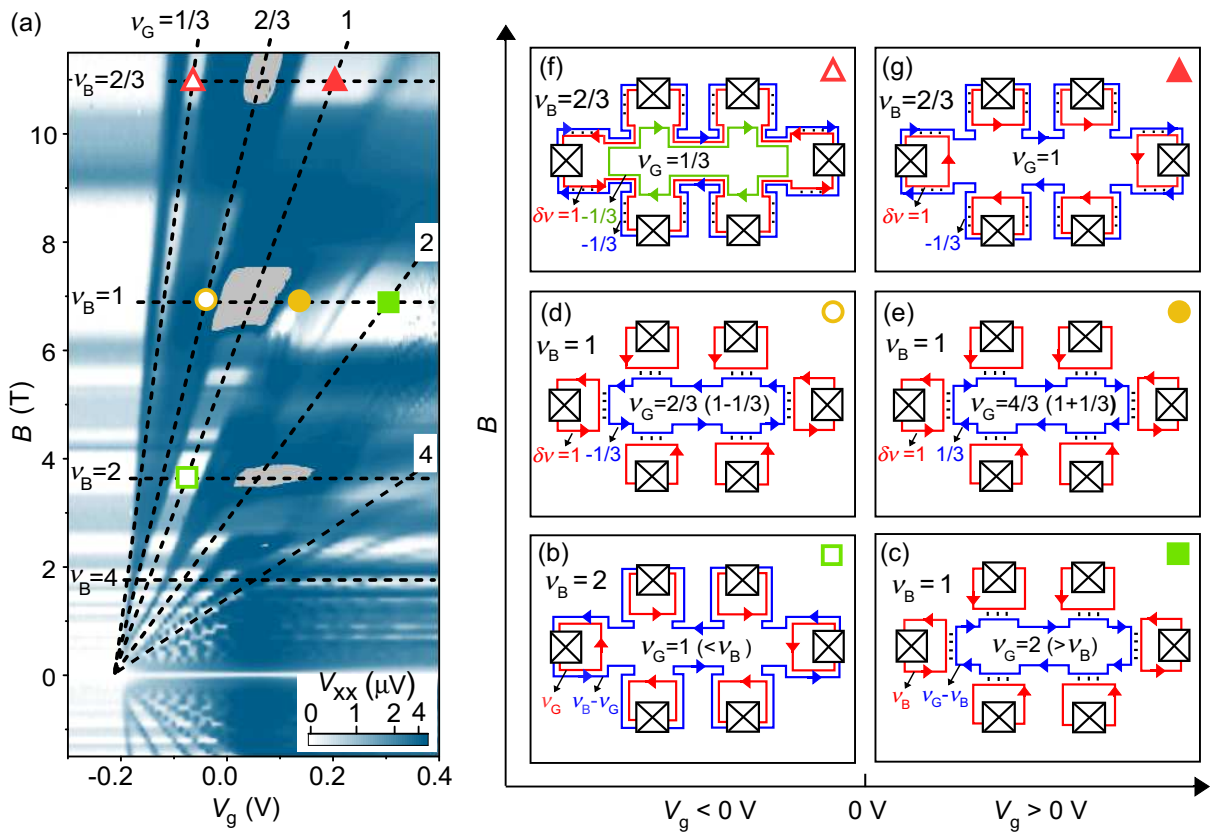}
\caption{(color online). (a) Color plot of $V_{\mathrm{xx}}$ measured as a function of gate voltage $V_{\mathrm{g}}$ and magnetic field $B$ for the $\ell$ = 100 $\mu$m device. Dashed lines indicate the filling factor $\nu_{\mathrm{B}}$ for the ungated region and $\nu_{\mathrm{G}}$ for the gated region. The data around $\nu_{\mathrm{B}}$ = $\nu_{\mathrm{G}}$ showing too small $G < 0.1 e^2/h$ to measure $V_{\mathrm{xx}}$ are grayed out. Representative edge channel structures at several conditions marked by open or filled symbols are sketched in (b-g). (b) Edge structure at $\nu_{\mathrm{G}}$ = 1 and $\nu_{\mathrm{B}}$ = 2, where the transport is dominated by the spin-down edge channel from spin-resolved second lowest Landau level. (c) Edge structure at $\nu_{\mathrm{G}}$ = 2 and $\nu_{\mathrm{B}}$ = 1, where closed channel formed underneath the gate couple to the outer channel through edge equilibration, marked by short bars. (d) Counter-propagating $\delta \nu$ = 1 and $\delta \nu$ = -1/3, and (e) co-propagating $\delta \nu$ = 1 and $\delta \nu$ = 1/3 edge channels are formed for equilibration at $\nu_{\mathrm{G}}$ = 2/3 and $\nu_{\mathrm{G}}$ = 4/3, respectively, in $\nu_{\mathrm{B}}$ = 1. (f) Edge structure at $\nu_{\mathrm{G}}$ = 1/3 and $\nu_{\mathrm{B}}$ = 2/3, where complicated edge equilibration between two $\delta \nu$ = -1/3 and one $\delta \nu$ = 1 channels is involved. (g) Edge structure at $\nu_{\mathrm{G}}$ = 1 and $\nu_{\mathrm{B}}$ = 2/3, where a sole $\delta \nu$ = -1/3 hole channel from 2/3 state contributes to the transport.}
\end{figure*}

In general, edge equilibration between inner and outer edge channels, $C_{\mathrm{i}}$ and $C_{\mathrm{o}}$ respectively, can be studied by preparing a scattering region of length $\ell$, as schematically shown in Fig.\,1(a) for a co-propagating configuration and Fig.\,1(b) for a counter-propagating one. The two channels must be separated in other regions, for example by spatially modulating the carrier density as shown (yellow region). This allows us to supply independent voltages ($V_{\mathrm{o/i}}^{\mathrm{S}}$) and measure the outcomes ($V_{\mathrm{o/i}}^{\mathrm{M}}$). Here we note that bulk equilibration might appear between the right- and left-going parts of channel $C_{\mathrm{i}}$, which influences the evaluation of the edge equilibration. This bulk equilibration must be taken into account in the estimation of edge equilibration, otherwise underestimates the edge equilibration. Scattering between $C_{\mathrm{i}}$ and $C_{\mathrm{o}}$ in the upper and lower regions should be negligible as compared to the main issue of edge equilibration with the shortest inter-channel distance. Scattering is absent on the left side of $C_{\mathrm{o}}$, if no other channels are formed. We propose and demonstrate a generalized Hall bar for evaluating both edge and bulk equilibration.

A generalized Hall bar is defined as shown in Fig.\,1(c), where an inner quantum Hall state at filling factor $\nu_{\mathrm{G}}$ is formed inside the outer quantum Hall state at $\nu_{\mathrm{B}}$ by applying gate voltage $V_{\mathrm{g}}$ on a gate shaped into a standard six-terminal Hall bar (yellow region). Both $\nu_{\mathrm{G}}$ and $\nu_{\mathrm{B}}$ can be tuned with perpendicular magnetic field $B$ and $V_{\mathrm{g}}$. If $\nu_{\mathrm{G}}$ and $\nu_{\mathrm{B}}$ are set at different QH states, as shown in the figure for $\nu_{\mathrm{G}}$ = 2 and $\nu_{\mathrm{B}}$ = 1, circulating edge channel $C_{\mathrm{i}}$ appears along the boundary between the inner and outer QH regions. Electrical connection to $C_{\mathrm{i}}$ can be made with Corbino-type contacts, each of which is formed by etching the heterostructure and patterning an ohmic contact to edge channel $C_{\mathrm{o}}$ along the perimeter of the etched region. With this edge configuration, the conductance between contacts $\Omega_1$ and $\Omega_6$ of the generalized Hall bar is sensitive to the edge equilibration between channels $C_{\mathrm{i}}$ and $C_{\mathrm{o}}$ in the IQH regime. The longitudinal voltage $V_{\mathrm{xx}}$ between contacts $\Omega_2$ and $\Omega_3$ is sensitive to the bulk equilibration for the inner QH state with the channel $C_{\mathrm{i}}$. Therefore, both edge and bulk equilibration can be studied with this device.

Previously, edge equilibration has been extensively studied on a single equilibration region by using a single cross gate \cite{Haug89,Haug90,Komiyama89,Komiyama92,Hirai95,Wurtz02}, double series gates \cite{Alphenaar90,Wees89,Kouwenhoven90,Muller90,Machida01,Roddaro09}, and non-ideal electric contacts \cite{Faist91a}, all of which rely on the absence of scattering in the bulk region. This may be sufficient for studying IQH states, but may not be suitable for FQH states with finite bulk scattering.

We have fabricated such generalized Hall bar devices on a standard GaAs/AlGaAs heterostructure with a two-dimensional electron gas (2DEG) located at 110 nm below the surface having electron density 1.7$\times $10$^{11}$ cm$^{-2}$ and low-temperature mobility 460 m$^2$/V\,s. The generalized Hall bar is designed with width $\ell$ = 10, 50, and 100 $\mu$m, and total length $L_t$ = 510 $\mu$m, as shown in an optical micrograph of Fig.\,1(d) for an $\ell$ = 100 $\mu$m device. The effective length for the bulk equilibration is $L_{\mathrm{eff}}$ = $L_1+L_2+L_3$ = $L_t$ - 2$\ell$. All ohmic contacts have a resistance of around 200 $\Omega$. We employed a constant voltage drive, where a low-frequency (37 Hz) AC voltage of amplitude $V_{\mathrm{AC}}$ = 30 $\mu$V was applied between the source contact $\Omega_1$ and the drain contact $\Omega_6$, as shown in Fig.\,1(c). The two-terminal conductance $G$ = $I_{\mathrm{AC}}/V_{\mathrm{AC}}$ as well as the longitudinal and transverse voltages, $V_{\mathrm{xx}}$ and $V_{\mathrm{xy}}$ respectively, were measured using standard lock-in technique. The measurements were taken in a dilution refrigerator at base temperature of 20 mK and in a magnetic field up to 12 T.

Figure 2(a) shows a color plot of $V_{\mathrm{xx}}$ measured as a function of $V_{\mathrm{g}}$ and $B$ for the $\ell$ = 100 $\mu$m device. Vanishing $V_{\mathrm{xx}}$ (white region) seen in the high field suggests negligible bulk equilibration in both inner and outer QH states. The observed pattern in $V_{\mathrm{xx}}$ can be understood with variations of $\nu_{\mathrm{B}}$ shown by the horizontal dash lines and $\nu_{\mathrm{G}}$ shown by inclined dash lines. In some regions near $V_{\mathrm{g}}$ = 0 V, where $\nu_{\mathrm{B}}$ = $\nu_{\mathrm{G}}$, $G$ was too small $(< 0.1 e^2/h)$ and the color plot is grayed out. In these regions, a QH region with uniform filling forms over the entire 2DEG, so that there is no edge channel connecting different ohmic contacts. At large negative gate voltage $V_{\mathrm{g}} < $ -0.21 V where the gated region is completely depleted with $\nu_{\mathrm{G}}$ = 0, the system reduces to a conventional anti-Hall bar with no electrons inside \cite{Mani94,Mani95,Mani97}. For $V_{\mathrm{g}} > $ 0 V, on the other hand, $V_{\mathrm{g}}$ of up to 0.4 V can be applied without a measurable gate leakage, where $\nu_{\mathrm{G}}$ reaches more than twice that of $\nu_{\mathrm{B}}$.

Representative channel configurations based on the hierarchical edge structure are illustrated in Figs.\,2(b-g). The simplest case for IQH states is shown in Fig.\,2(b) for $\nu_{\mathrm{G}}$ = 1 and $\nu_{\mathrm{B}}$ = 2, where the transport is dominated by the edge channel for spin-down electrons from spin-resolved second lowest Landau level (LL). Vanishing $V_{\mathrm{xx}}$ at this condition marked by the open square in Fig.\,2(a) and quantized Hall conductance $-e^2/h$ (not shown) ensure no bulk equilibration for $\nu_{\mathrm{G}}$ = 1 and $\nu_{\mathrm{B}}$ = 2. Another configuration shown in Fig.\,2(c) for $\nu_{\mathrm{G}}$ = 2 and $\nu_{\mathrm{B}}$ = 1 involves a closed loop of the inner edge channel from second lowest LL in the shape of a Hall bar, which is attached with edge equilibration to the outer edge channel that is connected to each ohmic contact from lowest LL, as seen in the data marked by the filled square in Fig.\,2(a). Similar configurations can be seen at any integer $\nu_{\mathrm{G}}$ greater than integer $\nu_{\mathrm{B}}$, where we can investigate both edge and bulk equilibration for IQH regimes as will be discussed in Sec. IV-A, with the model described in Sec. III.

Edge equilibration for a FQH state can be studied by defining the FQH state with $\nu_{\mathrm{G}}$ inside the host IQH state at $\nu_{\mathrm{B}}$. Figure 2(d) shows the configuration for studying edge equilibration between counter-propagating $\delta \nu$ = 1 and $\delta \nu$ = -1/3 channels at $\nu_{\mathrm{G}}$ = 2/3 and $\nu_{\mathrm{B}}$ = 1. This can be compared to the edge equilibration between co-propagating $\delta \nu$ = 1 and $\delta \nu$ = 1/3 channels at $\nu_{\mathrm{G}}$ = 4/3 and $\nu_{\mathrm{B}}$ = 1, as illustrated in Fig.\,2(e). Conveniently, the comparison can be made at a same $B$ as shown by the conditions marked by the open and filled circles in Fig.\,2(a). We study such equilibration in Sec. IV-B.

Edge equilibration may appear in other regions. Figure 2(g) shows the configuration at $\nu_{\mathrm{G}}$ = 1 and $\nu_{\mathrm{B}}$ = 2/3, where counter-propagating channels are equilibrated around the ohmic contact. Note that this is totally different from the situation in Fig.\,2(b) where the two co-propagating channels are already equilibrated as they come out from the same ohmic contact. Nevertheless, measurement in Fig.\,2(g) provides transport through a hole-conjugate edge channel. We observed vanishing $V_{\mathrm{xx}}$ at this condition marked by the filled triangle in Fig.\,2(a) and clear quantized Hall conductance of (1/3)$e^2/h$ (not shown), which indicates full edge equilibration and negligible bulk equilibration. This clearly demonstrates the reality of an isolated fractional hole edge channel of the $\nu_{\mathrm{B}}$ = 2/3 state nested in $\nu_{\mathrm{G}}$ = 1. This isolated channel is discussed in Sec. IV-C.

More complicated edge equilibration can be studied with $\nu_{\mathrm{G}}$ = 1/3 and $\nu_{\mathrm{B}}$ = 2/3 at the open triangle mark in Fig.\,2(a), which is shown in Fig.\,2(f). There should be three channels, two fractional hole channels and an integer channel, at the boundary of the QH states. This is also discussed in Sec. IV-C.

\begin{figure}[t]
\includegraphics[width=8.5 cm]{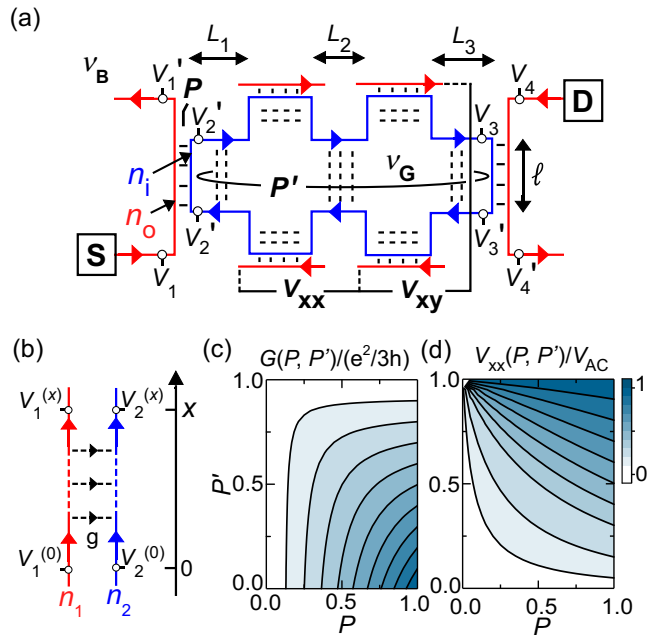}
\caption{(color online). A model for edge and bulk equilibration. (a) Edge channels in a generalized Hall bar, where edge and bulk equilibration are marked by dashed lines. The edge and bulk equilibrations are characterized by $P$ and $P'$, respectively. (b) Parallel channels with conductance $n_{\mathrm{1}}e^2/h$ and $n_{\mathrm{2}}e^2/h$ along $x$ axis. Uniform scattering conductivity $ge^2/h$ per unit length can be used to relate voltages $V_{1/2}^{(0/x)}$ along the channel. The calculated color plots of (c) conductance $G(P, P')$ normalized by $e^2/3h$ between source (S) and drain (D) contacts and (d) voltage $V_{\mathrm{xx}}$ normalized by source voltage $V_{\mathrm{AC}}$ are shown as a function of edge equilibration rate $P$ and bulk equilibration rate $P'$ for  $n_{\mathrm{o}}$ = 1  and $n_{\mathrm{i}}$ = -1/3.}
\end{figure}

\section{Modeling edge and bulk equilibration}

Before going to the data analysis, we present a model for edge and bulk equilibration in the generalized Hall bar. We consider fully incoherent regime by neglecting interference effects \cite{Sen08,Protopopov17,Nosiglia18}, and provide a way to evaluate the degree of equilibration from the measurement of $G$, $V_{\mathrm{xx}}$ and $V_{\mathrm{xy}}$. Figure 3(a) shows a typical edge channel configuration, as we see in Figs.\,2(c), 2(d), and 2(e). Here, the outer and inner edge channels have Hall conductance $n_{\mathrm{o}}e^2/h$ and $n_{\mathrm{i}}e^2/h$, respectively, where $n_{\mathrm{o}}$ and $n_{\mathrm{i}}$ can either be integer (1, 2, 3, ...), fraction (1/3) or a negative fraction (-1/3) for a hole-conjugate state. Following argument can be applied to both co- and counter-propagating channels with this definition, and can be adapted to other configurations seen in Figs.\,2(b), 2(f), and 2(g).

Scattering between two parallel channels with conductance $n_{\mathrm{1}}e^2/h$ and $n_{\mathrm{2}}e^2/h$ can be modeled with inter channel scattering conductance of $ge^2/h$ per unit length, as shown in Fig.\,3(b). The voltages $V_1^{(x)}$ and $V_2^{(x)}$ of the two channels change along $x$ axis with
\begin{equation}
\frac{d}{dx} \left( \begin{array}{c} V_{\mathrm{1}} \\ V_{\mathrm{2}} \end{array} \right) \mbox{~=~} \left( \begin{array}{cc} -g/n_{\mathrm{1}} & g/n_{\mathrm{1}}  \\ g/n_\mathrm{2} & -g/n_\mathrm{2} \end{array} \right) \left( \begin{array}{c} V_{\mathrm{1}} \\ V_{\mathrm{2}} \end{array} \right)
\end{equation}
under the current conservation. This provides a relation between voltages $V_{1/2}^{(0/x)}$ of channels 1 and 2 at two locations separated by $x$.

For $n_{\mathrm{1}} + n_{\mathrm{2}} \neq 0$ ($n_{\mathrm{1}}$ and/or $n_{\mathrm{2}}$ can be negative), we obtained the relation
\begin{equation}
\left( \begin{array}{c} V_{\mathrm{1}}^{(x)} \\ V_{\mathrm{2}}^{(x)} \end{array} \right) \mbox{~=~} \left( \begin{array}{cc} 1-\eta_2\xi & \eta_2\xi \\ \eta_1\xi & 1-\eta_1\xi \end{array} \right) \left( \begin{array}{c} V_{\mathrm{1}}^{\mathrm{(0)}} \\ V_{\mathrm{2}}^{\mathrm{(0)}} \end{array} \right)
\end{equation}
where we defined $\eta_1 = n_{\mathrm{1}}/(n_{\mathrm{1}} + n_{\mathrm{2}})$, $\eta_2 = n_{\mathrm{2}}/(n_{\mathrm{1}} + n_{\mathrm{2}})$, and $\xi = 1 - e^{-\lambda x}$ with $\lambda = g/n_{\mathrm{1}} + g/n_{\mathrm{2}}$. This describes the voltage change in the equilibration. As $\lambda$ as well as $\xi$ can be positive or negative depending on the signs of $n_{\mathrm{1}}$ and $n_{\mathrm{2}}$, it is convenient to introduce equilibration rate
\begin{equation}
P = 1 - e^{-|\lambda|x}
\end{equation}						
which changes from 0 for no equilibration ($x$ = 0) to 1 for full equilibration ($x$ = $\infty$). These relations are used to characterize the edge equilibration in Fig.\,3(a) with $n_{\mathrm{o}}$ = $n_1$ and $n_{\mathrm{i}}$ = $n_2$.

The solution for $n_{\mathrm{1}} + n_{\mathrm{2}} = 0$ can be written as
\begin{equation}
\left( \begin{array}{c} V_{\mathrm{1}}^{(x)} \\ V_{\mathrm{2}}^{(x)} \end{array} \right) \mbox{~=~} \frac{1}{1-\xi'} \left( \begin{array}{cc} 1-2\xi' & \xi' \\ -\xi' & 1 \end{array} \right) \left( \begin{array}{c} V_{\mathrm{1}}^{\mathrm{(0)}} \\ V_{\mathrm{2}}^{\mathrm{(0)}} \end{array} \right)
\end{equation}
with $\xi' = g/(g + n_{\mathrm{1}}/x)$, where $n_1$ (= $-n_2$) can be positive or negative. We use this relation for characterizing bulk equilibration between counter-propagating $n_{\mathrm{i}}$ (= $n_1$) and $-n_{\mathrm{i}}$ (= $n_2$) channels in Fig.\,3(a) by introducing bulk equilibration rate
\begin{equation}
P' = g/(g + |n_{\mathrm{i}}|/x)
\end{equation}
which also changes from 0 for no bulk equilibration ($x$ = 0) to 1 for full bulk equilibration ($x$ = $\infty$).

Equations (2) and (4) are used to relate the voltages at specific points, $V_1$, $V_1'$, $V_2$, etc., in Fig.\,3(a). The edge equilibration is assumed to be identical for all six terminals. For the bulk equilibration, scattering in the arms of $C_{\mathrm{i}}$ for the voltage probes is effectively absent, as far as no current flows through the voltage probes. Then, the bulk equilibration takes place in the main channel of effective length $L_{\mathrm{eff}} = L_t - 2\ell$. With equilibration rates $P$ and $P'$, we obtained two-terminal conductance $G$, longitudinal voltage $V_{\mathrm{xx}}$, and transverse voltage $V_{\mathrm{xy}}$ as
\begin{subequations}\label{grp}
\begin{align}
G(P, P') &= \frac{e^2}{h} \frac{n_{\mathrm{o}} n_{\mathrm{i}} \zeta(1-\zeta')}{n_{\mathrm{o}}+n_{\mathrm{i}}+(1-2\zeta')[n_{\mathrm{o}}(1-\zeta)+n_{\mathrm{i}}]} \label{second} \\
V_{\mathrm{xx}}(P, P') &= \frac{n_{\mathrm{o}} \zeta \zeta' V_{\mathrm{AC}}}{n_{\mathrm{o}}+n_{\mathrm{i}}+(1-2\zeta')[n_{\mathrm{o}}(1-\zeta)+n_{\mathrm{i}}]} \frac{L_2}{L_{\mathrm{eff}}} \label{third} \\
V_{\mathrm{xy}}(P, P') &= \frac{n_{\mathrm{o}} \zeta(1-\zeta')V_{\mathrm{AC}}}{n_{\mathrm{o}}+n_{\mathrm{i}}+(1-2\zeta')[n_{\mathrm{o}}(1-\zeta)+n_{\mathrm{i}}]} \label{fourth}
\end{align}
\end{subequations}
where $\zeta$ is a function of $P$ with $\zeta = P$ for $\lambda > 0$ and $\zeta = -P/(1-P)$ for $\lambda < 0$, and $\zeta'$ is a function of $P'$ with $\zeta' = P'$ for $n_{\mathrm{i}} > 0$ and $\zeta' = -P'/(1-2P')$ for $n_{\mathrm{i}} < 0$.

We used these equations to evaluate $P$ and $P'$ from the measured values. While $G$ and $V_{\mathrm{xx}}$ are used in the following evaluation, consistency with $V_{\mathrm{xy}}$ is also confirmed. Figure 3(c) and 3(d) shows how $G(P, P')$ and $V_{\mathrm{xx}}(P, P')$, respectively, change with $P$ and $P'$ for $n_{\mathrm{o}}$ = 1 and $n_{\mathrm{i}}$ = -1/3, which is the case of $\nu_{\mathrm{G}}$ = 2/3 and $\nu_{\mathrm{B}}$ = 1 in Fig.\,2(d). In the evaluation of edge equilibration for this case, a large sample with $P$ = 1 and $P'$ = 0 exhibits maximum conductance $G_{max} =(1/3)(e^2/h)$. For a small sample having a short $\ell$, the conductance decreases with either reduction of edge equilibration ($P < 1$) or enhancement of bulk equilibration ($P' > 0$). Therefore, one has to evaluate both $P$ and $P'$. This is particularly important in evaluation of short equilibration length in the hole-conjugate state.

\section{Analysis and Discussions}
\subsection{Integer QH regime}

We start from equilibration in the IQH regime with $\nu_{\mathrm{B}}$ = 4 at $B$ = 1.7 T to test the scheme. Figures 4(a-c) show the gate voltage $V_{\mathrm{g}}$ dependence of $G$, $V_{\mathrm{xx}}$, and $V_{\mathrm{xy}}$ for the $\ell$ = 100 $\mu$m device. Corresponding $\nu_{\mathrm{G}}$ is shown in the top scale. While more than two edge channels are involved in the system, we use the two-channel model of Eqs.\,(6a-6c) by regarding the system as composed of two bundles of edge channels acting as two channels with conductance $n_{\mathrm{o}}e^2/h$ and $n_{\mathrm{i}}e^2/h$ with $n_{\mathrm{o}}$, $n_{\mathrm{i}} >$ 1.

For $V_{\mathrm{g}} <$ 0 V with $\nu_{\mathrm{G}} < \nu_{\mathrm{B}}$, quantized transport is clearly seen as a series of conductance plateaus in Fig.\,4(a), vanishing $V_{\mathrm{xx}}$ in Fig.\,4(b), and maximum $|V_{\mathrm{xy}}|$ ($\simeq V_{\mathrm{AC}}$) in Fig.\,4(c) under our constant voltage ($V_{\mathrm{AC}}$) drive. Negative $V_{\mathrm{xy}}$ is associated with the lower electron density underneath the gate region as compared to the outside bulk region. Here, a bundle of $n_{\mathrm{i}}$ (= $\nu_{\mathrm{B}} - \nu_{\mathrm{G}}$) channels from energetically higher-lying Landau levels contributes the transport, while the other bundle of $n_{\mathrm{o}}$ (= $\nu_{\mathrm{G}}$) channels from lower-lying Landau levels are fully reflected to the ohmic contact as seen in Fig.\,2(b). As the two bundles coming out from the same ohmic contact are always equilibrated, we can evaluate bulk equilibration rate $P'$ only, which is plotted in Fig.\,4(e).

\begin{figure}[t]
\includegraphics[width=8.5 cm]{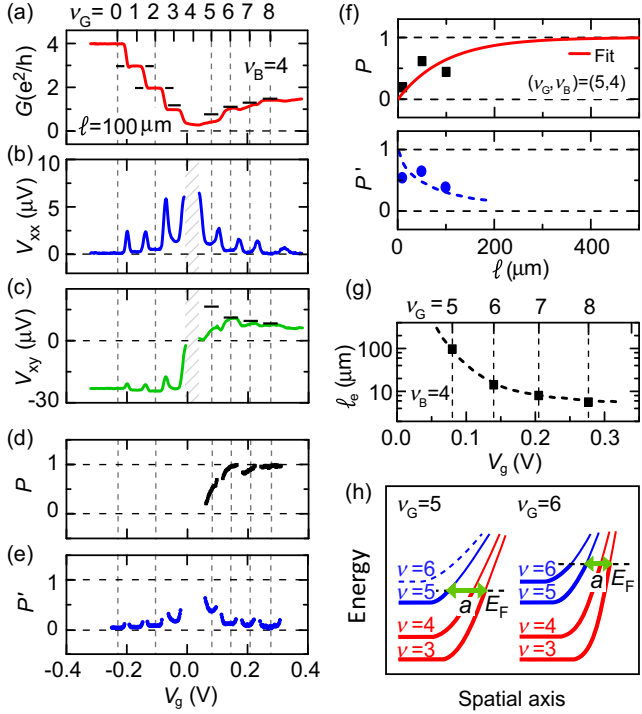}
\caption{(color online).  Gate voltage $V_{\mathrm{g}}$ dependence of (a) $G$, (b) $V_{\mathrm{xx}}$, (c) $V_{\mathrm{xy}}$ and (d) $P$, (e) $P'$ at $\nu_{\mathrm{B}}$ = 4 for the $\ell$ = 100 $\mu$m device. Corresponding filling factor $\nu_{\mathrm{G}}$ is shown in the top scale. Reference levels with $P$ = 1 and $P'$ = 0 are marked as horizontal bars. (f) $\ell$ dependence of $P$ and $P'$ at $\nu_{\mathrm{G}}$ = 5 and $\nu_{\mathrm{B}}$ = 4. The solid red line in $P$ shows a fit to Eq.\,(3) for extracting the equilibration length $\ell_{\mathrm{e}}$. The blue dash line in $P'$ is guided for the eye. (g) $V_{\mathrm{g}}$ dependence of $\ell_{\mathrm{e}}$. (h) Landau levels gained by the edge potential for $\nu_{\mathrm{G}}$ = 5 and $\nu_{\mathrm{B}}$ = 4 in the left panel, and $\nu_{\mathrm{G}}$ = 6 and $\nu_{\mathrm{B}}$ = 4 in the right panel. The distance $a$ between the outer bundle with $n_{\mathrm{o}}$ (= $\nu_{\mathrm{B}}$) and the inner bundle with $n_{\mathrm{i}}$ (= $\nu_{\mathrm{G}}$ - $\nu_{\mathrm{B}}$) is shorter for larger $\nu_{\mathrm{G}}$.}
\end{figure}

Similar quantized transport is also seen at $V_{\mathrm{g}} >$ 0 V with $\nu_{\mathrm{G}} > \nu_{\mathrm{B}}$, where plateaus in $G$, vanishing $V_{\mathrm{xx}}$, and positive plateaus in $V_{\mathrm{xy}}$ are resolved. In this situation, the edge and bulk equilibration can be evaluated for the two bundles of $n_{\mathrm{o}}$ ($\simeq$ $\nu_{\mathrm{B}}$) and $n_{\mathrm{i}}$ ($\simeq$ $\nu_{\mathrm{G}} - \nu_{\mathrm{B}}$) channels as seen in Fig.\,2(c). Finite $G$ indicates the presence of edge equilibration, and non-zero $V_{\mathrm{xx}}$ shows the presence of bulk equilibration. As a reference, maximum $G$ and $V_{\mathrm{xy}}$ values for $P$ = 1 and $P'$ = 0 are shown by horizontal bars in Fig.\,4(a) and 4(c). The deviations from these values should be discussed with $P$ and $P'$, which are obtained by using Eqs.\,(6a-6c) as shown in Figs.\,4(d) and 4(e), respectively.

The bulk equilibration $P'$ is minimized at integer $\nu_{\mathrm{G}}$ with a well-defined IQH state in the gated region, and increases significantly at non-integer $\nu_{\mathrm{G}}$ with a conductive bulk region. The minimum $P'$ values at $\nu_{\mathrm{G}}$ = 3 and 5 are slightly higher than those at $\nu_{\mathrm{G}} \leq$ 2 and $\nu_{\mathrm{G}} \geq$ 6, which might be reasoned to the small Zeeman gap of $\nu_{\mathrm{G}}$ = 3 and 5 IQH states.

The edge equilibration in the IQH regime can be understood with the edge potential profile, as shown in Fig.\,4(h) for $V_{\mathrm{g}} >$ 0 V. Spin-conserving scattering between $\nu$ = 5 channel and $\nu$ = 3 channel should dominate the edge equilibration. Since the scattering rate should increase with decreasing the spatial distance, $a$, between the two channels, larger $P$ is expected with smaller $a$. In our situation, this distance decreases with increasing $V_{\mathrm{g}}$ as the edge potential becomes steeper with increasing gate induced charge. This explains the observed increase of $P$ with $V_{\mathrm{g}}$ in Fig.\,4(d).

\begin{figure}[t]
\includegraphics[width=8.5 cm]{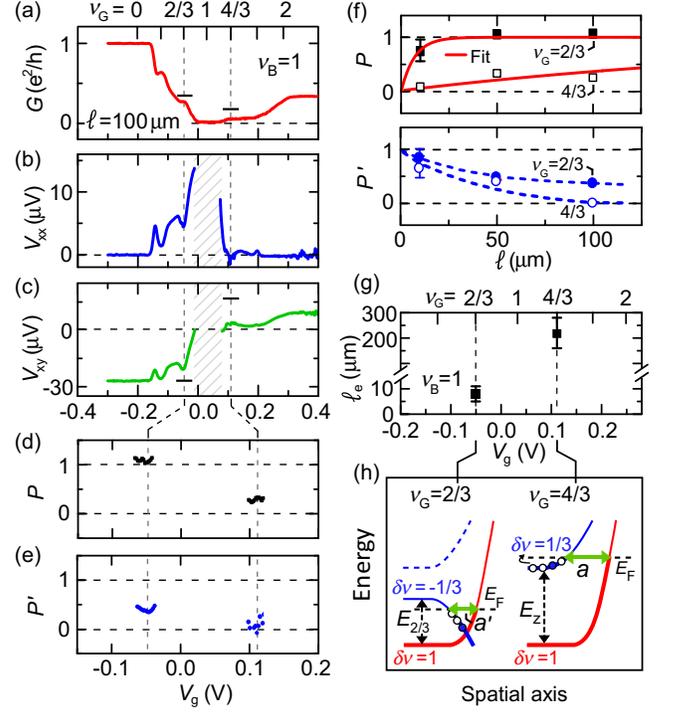}
\caption{(color online). Gate voltage $V_{\mathrm{g}}$ dependence of (a) $G$, (b) $V_{\mathrm{xx}}$, (c) $V_{\mathrm{xy}}$ and (d) $P$, (e) $P'$ at $\nu_{\mathrm{B}}$ = 1 for the $\ell$ = 100 $\mu$m device. Corresponding filling factor $\nu_{\mathrm{G}}$ is shown in the top scale. Reference levels with $P$ = 1 and $P'$ = 0 are marked as horizontal bars for comparison with the experiment data. (f) $\ell$ dependence of $P$ and $P'$ at $\nu_{\mathrm{G}}$ = 2/3 and $\nu_{\mathrm{G}}$ = 4/3. The solid red lines in $P$ are the fit to Eq.\,(2) for extracting the equilibration length $\ell_{\mathrm{e}}$. The blue dash lines in $P'$ are guided for the eye. (g) $V_{\mathrm{g}}$ dependence of $\ell_{\mathrm{e}}$. (h) The band structure for $\nu_{\mathrm{G}}$ = 2/3 (left) and $\nu_{\mathrm{G}}$ = 4/3 (right) formed in a same magnetic field. At $\nu_{\mathrm{G}}$ = 2/3, the counter-propagating channels for edge equilibration are originated from the up bending $\delta \nu$ = 1 band and the down bending $\delta \nu$ = -1/3 band with a band gap of $E_{2/3}$. While at $\nu_{\mathrm{G}}$ = 4/3, the co-propagating channels for edge equilibration are originated from both the up bending spin-up band with $\delta \nu$ = 1 and the spin-down band with $\delta \nu$ = 1/3. The two bands are separated by Zeeman energy $E_Z$. The inter-channel distance is labeled as $a'$ for $\nu_{\mathrm{G}}$ = 2/3 and $a$ for $\nu_{\mathrm{G}}$ = 4/3.}
\end{figure}

Similar results were obtained with other devices of $\ell$ = 50 and 10 $\mu$m. Figure 4(f) shows the $\ell$ dependence of $P$ and $P'$ for $\nu_{\mathrm{G}}$ = 5 and $\nu_{\mathrm{B}}$ = 4. While the data points are scattered possibly with the sample-specific characteristics, the edge equilibration length $\ell_{\mathrm{e}}$ defined as $1/|\lambda|$ is estimated to be about 100 $\mu$m by fitting $P$ to Eq.\,(3) \cite{Note_IQH}. This $\ell_{\mathrm{e}}$ estimated from several devices decreases with increasing $\nu_{\mathrm{G}}$, as shown in Fig.\,4(g). It should be noted that $P'$ significantly increases with decreasing $\ell$ in Fig.\,4(f). If the bulk equilibration were not considered in the analysis, the measured $G$ would have been analyzed with $P'$ = 0 in Eqs.\,(6a-6c). This overestimates $\ell_{\mathrm{e}}$ to $\sim$ 140 $\mu$m, 40 \% higher than the data in Fig.\,4(f). This clearly demonstrates the importance of evaluating both $P$ and $P'$, which could be more crucial for the following FQH regimes.

\subsection{Fractional QH regime}

Next we study equilibration at $\nu_{\mathrm{B}}$ = 1 at $B$ = 6.5 T, where we can study the edge equilibration between counter-propagating $\delta \nu$ = 1 and $\delta \nu$ = -1/3 channels for hole-conjugate fractional state $\nu_{\mathrm{G}}$ = 2/3, and between co-propagating $\delta \nu$ = 1 and $\delta \nu$ = 1/3 channels for particle fractional state $\nu_{\mathrm{G}}$ = 4/3, as seen in the edge channel structures in Figs.\,2(d) and 2(e), respectively. More complicated channel structure can appear when the edge potential is smooth \cite{Meir94,Sabo17}. Since we have not tested the actual structure in our device, the original and simplest model with $\delta \nu$ = 1 and $\delta \nu$ = -1/3 channels is applied in the following analysis. Figures 5(a-c) show the gate voltage $V_{\mathrm{g}}$ dependence of $G$, $V_{\mathrm{xx}}$, and $V_{\mathrm{xy}}$ for the $\ell$ = 100 $\mu$m device. Presence of fractional states is visible with conductance steps in Fig.\,5(a), dips in $V_{\mathrm{xx}}$ of Fig.\,5(b) and also in $V_{\mathrm{xy}}$ of Fig.\,5(c). Both edge and bulk equilibration is obvious as the data in $G$ and $V_{\mathrm{xy}}$ is greatly deviated from the reference level for $P$ = 1 and $P'$ = 0 shown by horizontal bars. Evaluated $P$ and $P'$ with Eqs.\,(6a-6c) are plotted in Figs.\,5(d) and 5(e), respectively. We find notable difference in equilibration between $\nu_{\mathrm{G}}$ = 2/3 and $\nu_{\mathrm{G}}$ = 4/3. While the small $G$ at $\nu_{\mathrm{G}}$ = 4/3 is mainly associated with a small edge equilibration rate $P$ $\sim$ 0.3 ($P'$ $\sim$ 0), the small $G$ at $\nu_{\mathrm{G}}$ = 2/3 is mainly associated with a large bulk equilibration rate $P'$ $\sim$ 0.5 ($P$ $\sim$ 1).

The bulk equilibration can be understood with excitation to higher lying states. In the composite fermion (CF) picture, $\nu$ = 2/3 state corresponds to filling factor $\nu^{\mathrm{CF}}$ = -2 for composite fermions, and $\nu$ = 4/3 state is its particle-hole counterpart corresponding to a same filling factor. While this particle-hole symmetry might imply that these two states have a same activation energy, the symmetry can be practically broken in the presence of Landau level mixing or else, as reported in experiments and theories \cite{Kukushkin99,Liu14,Padmanabhan10,Zhang16,Chang83}. In our measurement, the electron density is different for $\nu_{\mathrm{G}}$ = 2/3 and 4/3 states, and this may cause different disorder effects. Such asymmetry could be the reason for the difference of small $P'$ for $\nu_{\mathrm{G}}$ = 4/3 and large $P'$ for $\nu_{\mathrm{G}}$ = 2/3 in our device, as shown in Fig.\,5(e). Systematic study together with heat transport through the bulk \cite{Altimiras12} may be useful in understanding the asymmetry.

In contrast, the edge equilibration can be understood with the spatial distribution of the edge channels. For $\nu_{\mathrm{G}}$ = 4/3, co-propagating integer (1) and fractional (1/3) channels are separated by $a$, as shown in the right panel of Fig.\,5(h), which is determined by the electrostatic edge potential and the energy gap between the integer and fractional levels \cite{Deviatov08}, analogous to the IQH regime \cite{Wurtz02}. However, the edge structure proposed by MacDonald \cite{Macdonal90} for $\nu_{\mathrm{G}}$ = 2/3 is composed of an up bending $\delta \nu$ = 1 band and a down bending $\delta \nu$ = -1/3 band, as shown in the left panel of Fig.\,5(h). The inter-channel distance $a'$ is determined by the interaction \cite{Sarma97,Chang03} and may be affected by the edge potential and the energy gap \cite{Meier14}. Therefore, $a'$ can be very short comparable to the magnetic length \cite{Wen92,Meir94} and shorter than $a$ for $\nu_{\mathrm{G}}$ = 4/3. This explains the distinct difference in the edge equilibration rate $P$, close to 1 at $\nu_{\mathrm{G}}$ = 2/3 but $\sim$ 0.3 at $\nu_{\mathrm{G}}$ = 4/3 in Fig.\,5(d). As shown in the $\ell$ dependence of $P$ in Fig.\,5(f), the edge equilibration length $\ell_{\mathrm{e}}$ can be determined by fitting $P$ to Eq.\,(3). As summarized in Fig.\,5(g), $\ell_{\mathrm{e}}$ $\simeq$ 8 $\mu$m at $\nu_{\mathrm{G}}$ = 2/3, is much shorter than $\ell_{\mathrm{e}}$ $\simeq$ 200 $\mu$m at $\nu_{\mathrm{G}}$ = 4/3. The obtained $\ell_{\mathrm{e}}$ $\simeq$ 8 $\mu$m for 2/3 state does not contradict to the previous report in Ref.\cite{Grivnin14}.

\subsection{Isolated fractional hole channel}

Finally, we discuss fractional edge channel transport for the two configurations, (i) $\nu_{\mathrm{G}}$ = 1/3 and $\nu_{\mathrm{B}}$ = 2/3 and (ii) $\nu_{\mathrm{G}}$ = 1 and $\nu_{\mathrm{B}}$ = 2/3. The edge channel structures have been sketched in Figs.\,2(f) and 2(g). For both cases, the counter-propagating $\delta \nu$ = 1 and -1/3 channels emanating from the contact must be equilibrated before reaching the Hall bar region, as the counter-propagating length (280 $\mu$m) is much longer than the equilibration length ($\ell_{\mathrm{e}}$ $\simeq$ 8 $\mu$m). Figures 6(a-c) show $V_{\mathrm{g}}$ dependence of $G$, $V_{\mathrm{xx}}$, $V_{\mathrm{xy}}$ for the $\ell$ = 100 and 10 $\mu$m devices.

For $\nu_{\mathrm{G}}$ = 1/3 at $V_{\mathrm{g}} <$ 0 V, complicated equilibration takes place between an inner closed $\delta \nu$ = -1/3 channel and the outer two counter-propagating channels ($\delta \nu$ = 1 and -1/3) along the gate boundary. Since the outer $\delta \nu$ = -1/3 and $\delta \nu$ = 1 channels come from the same ohmic contact, we consider a bundle of these two channels with effective $\delta \nu$ = 2/3, edge equilibration rate $P$ between this bundle and the inner $\delta \nu$ = -1/3, and bulk equilibration rate $P'$ for $\nu_{\mathrm{G}}$ = 1/3 state. This modified model suggests $P \simeq$ 1 for all devices. Deviation in $G$ and $V_{\mathrm{xy}}$ from the ideal limit at $P'$ = 0 (horizontal bars) can be seen for $\ell$ = 10 $\mu$m, together with finite $V_{\mathrm{xx}}$, suggesting the presence of bulk equilibration for a small $\ell$.

\begin{figure}
\includegraphics[width=6 cm]{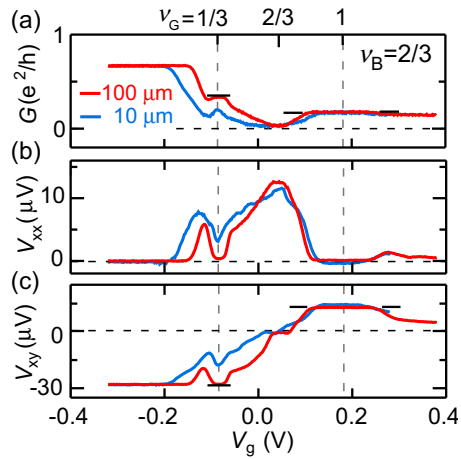}
\caption{(color online). Gate voltage $V_{\mathrm{g}}$ dependence of (a) $G$, (b) $V_{\mathrm{xx}}$, and (c) $V_{\mathrm{xy}}$ at $\nu_{\mathrm{B}}$ = 2/3 for the $\ell$ =100 and 10 $\mu$m devices. Corresponding filling factor $\nu_{\mathrm{G}}$ is shown in the top scale. Reference levels with $P$ =1 and $P'$ = 0 are marked as horizontal bars for comparison with the experiment data.}
\end{figure}

Similar analysis is made for $\nu_{\mathrm{G}}$ = 1 at $V_{\mathrm{g}}>$ 0 V, where $G$ and $V_{\mathrm{xy}}$ are consistent with no bulk equilibration ($P'$ = 0), as shown by the horizontal bars. This is manifested by vanishing $V_{\mathrm{xx}}$ even in the smallest sample with $\ell$ = 10 $\mu$m. This indicates that a clean isolated $\delta \nu$ = -1/3 channel without bulk equilibration is formed.
Here, the bulk equilibration is significantly suppressed as the high magnetic field ($B$ = 10.5 T) provides a large $\nu_{\mathrm{G}}$ = 1 gap \cite{note2} in the gated region and fractional $\nu_{\mathrm{B}}$ = 2/3 gap in the bulk [see Fig. 2(g)].

This clean fractional $\delta \nu$ = -1/3 edge channel between $\nu_{\mathrm{G}}$ = 1 and $\nu_{\mathrm{B}}$ = 2/3 may permit various experiments. For example, when it is weakly coupled to an integer edge channel of $\delta \nu$ = 1, one can study the edge equilibration as well as charge and neutral modes with various distances. This could be useful in further investigation of hole-conjugate fractional states.

\section{Conclusion}

In summary, we have studied the charge equilibration in both IQH and FQH edge channels using a generalized Hall bar, in which the multi terminal geometry allows us to clearly separate the edge and bulk equilibration in electron transport, making an access to the equilibration problem especially for hole-conjugate FQH states to be possible. Based on such separation, we first analyzed the edge and bulk equilibration in IQH regimes. The observed equilibration behaviors can be well explained with the changes of the inter channel separation. For FQH regime, the equilibration between counter-propagating $\delta \nu$ = 1 and $\delta \nu$ = -1/3 edge channels for hole-conjugate 2/3 state, and between co-propagating $\delta \nu$ = 1 and $\delta \nu$ = 1/3 edge channels for particle-like 4/3 state are studied. The characteristic equilibration length, $\ell_{\mathrm{e}}$ $\simeq$ 8 $\mu$m is quantitatively determined for this hole-conjugate 2/3 state, which is found to be much smaller than $\ell_{\mathrm{e}}$ $\simeq$ 200 $\mu$m for the particle 4/3 state. Furthermore, clean transport of a $\delta \nu$ = -1/3 hole channel in 2/3 state without showing bulk equilibration is identified at filling factors ($\nu_{\mathrm{G}}$ = 1, $\nu_{\mathrm{B}}$ = 2/3) with a positive gate bias.

As the edge equilibration in counter-propagating edge channels in hole-conjugate FQH states is associated with a completely different band structure comparing to the co-propagating case, which has rarely been explored, therefore, based on the generalized Hall bar scheme, it will be interesting to see whether such an unique band structure can be manifested in the equilibration behaviors by tuning the inter channel distance in magnetic fields and gate voltages, which would open a new avenue in studying and understanding those intriguing edge structures for the hole-conjugate FQH states.

\vspace{1cm}

\begin{acknowledgements} This work was supported by JSPS KAKENHI (JP15H05854 and JP26247051) and International Research Center for Nanoscience and Quantum Physics at Tokyo Institute of Technology.
\end{acknowledgements}


\begin{thebibliography}{99}
\bibitem{Note_IQH} In addition to the determined equilibration length, we also estimated the standard error as $\Delta \ell_e$ = 40 $\mu$m to take account the scattering of the points.
\bibitem{note2} The Fermi level in $\nu_{\mathrm{G}}$ = 1 region with a positive gate bias is located between spin-splitting bands, it is known that the splitting can be dramatically enhanced by electron-electron interactions, leading to a much larger Zeeman energy gap \cite{Englert82}.
\end{thebibliography}
\end{document}